\documentclass{article}

\usepackage{arxiv}
\usepackage{amsmath} 

\usepackage[utf8]{inputenc} 
\usepackage[T1]{fontenc}    
\usepackage{hyperref}       
\usepackage{url}            
\usepackage{booktabs}       
\usepackage{amsfonts}       
\usepackage{nicefrac}       
\usepackage{microtype}      
\usepackage{lipsum}		
\usepackage{graphicx}
\usepackage{natbib}
\usepackage{doi}
\usepackage{tikz}
\usetikzlibrary{shapes.geometric, arrows.meta, positioning} 

\usepackage[table]{xcolor}
\usepackage{colortbl}
\usepackage{adjustbox}

\definecolor{shade0}{RGB}{255,255,255}  
\definecolor{shade1}{RGB}{255,249,242}
\definecolor{shade2}{RGB}{255,244,230}
\definecolor{shade3}{RGB}{255,238,217}
\definecolor{shade4}{RGB}{255,232,204}
\definecolor{shade5}{RGB}{255,226,191}
\definecolor{shade6}{RGB}{255,221,179}
\definecolor{shade7}{RGB}{255,215,166}
\definecolor{shade8}{RGB}{255,209,153}
\definecolor{shade9}{RGB}{255,203,140}
\definecolor{shade10}{RGB}{255,198,128}
\definecolor{shade11}{RGB}{255,192,115}
\definecolor{shade12}{RGB}{255,186,102}
\definecolor{shade13}{RGB}{255,180,89}
\definecolor{shade14}{RGB}{255,175,77}
\definecolor{shade15}{RGB}{255,169,64}
\definecolor{shade16}{RGB}{255,163,51}
\definecolor{shade17}{RGB}{255,157,38}
\definecolor{shade18}{RGB}{255,152,26}
\definecolor{shade19}{RGB}{255,146,13}
\definecolor{shade20}{RGB}{255,140,0}   

\newcommand{\shadecell}[2]{\cellcolor{#1}{\scriptsize #2}}

\title{An Active Learning Framework for Data-Efficient, Human-in-the-Loop Enzyme Function Prediction}

\author{ \href{https://orcid.org/0000-0002-0991-7726}{\includegraphics[scale=0.06]{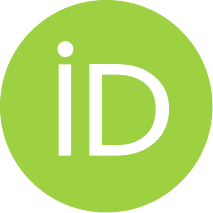}\hspace{1mm}Ashley Babjac} \\
	Department of Marine Sciences\\
	University of Georgia\\
	Athens, GA\\
	\texttt{Ashley.Babjac@uga.edu} \\
	\And
	\href{https://orcid.org/0000-0003-1006-9661}{\includegraphics[scale=0.06]{orcid.pdf}\hspace{1mm}Adrienne Hoarfrost} \\
	Department of Marine Sciences\\
	University of Georgia\\
	Athens, GA\\
	\texttt{Adrienne.Hoarfrost@uga.edu} \\
}

\renewcommand{\shorttitle}{HATTER | Human-in-the-loop Adaptive Toolkit for Transferable Enzyme Representations}

\hypersetup{
pdftitle={HATTER | Human-in-the-loop Adaptive Toolkit for Transferable Enzyme Representations},
pdfsubject={HITL Active Learning Software},
pdfauthor={Ashley Babjac and Adrienne Hoarfrost},
pdfkeywords={Active learning, prediction of enzyme functions, protein language modeling, experimental validation, generalization, human-in-the-loop},
}

\begin{document}
\maketitle

\begin{abstract}

Generalizable protein function prediction is increasingly constrained by the growing mismatch between exponentially expanding sequences of environmental proteins and the comparatively slow accumulation of experimentally verified functional data. Active learning offers a promising path forward for accelerating biological function prediction, by selecting the most informative proteins to experimentally annotate for data-efficient training, yet its potential remains largely unexplored. We introduce \textbf{HATTER} (Human-in-the-loop Adaptive Toolkit for Transferable Enzyme Representations), a modular framework that integrates multiple active learning strategies with human-in-the-loop experimental annotation to efficiently fine tune function prediction models. We compare active learning training to standard supervised training for biological enzyme function prediction, demonstrating that active learning achieves performance comparable to standard training across diverse protein sequence evaluation datasets while requiring fewer model updates, processing less data, and substantially reducing computational cost. Interestingly, point-based uncertainty sampling methods like entropy or margin sampling perform as well or better than more complex acquisition functions such as bayesian sampling or BALD, highlighting the relative importance of sequence diversity in training datasets and model architecture design. These results demonstrate that human-in-the-loop active learning can efficiently accelerate enzyme discovery, providing a flexible platform for adaptive, scalable, and expert-guided protein function prediction.

\end{abstract}



\section{Introduction}

The landscape of protein discovery is changing faster than our ability to understand it. Advancing sequencing technologies can now detect millions of novel proteins from uncultured microbes and environmental samples---or generate them with synthetic biology pipelines---yet only a minuscule fraction of these proteins have experimentally validated functions. The overwhelming majority fall into a growing ``unknown'' space: divergent sequences with no close homologs and uncertain function. Labeling these novel proteins with their molecular function relies on relatively time-consuming and costly experimental characterization. As this data gap widens, there is an increasing need for computational approaches that can efficiently leverage limited experimental annotations for function prediction on novel sequences. Intelligently choosing which proteins to characterize experimentally is therefore a central challenge in scaling functional annotation and training predictive models efficiently.

Deep learning models have made remarkable progress in enzyme function prediction, powering widely used systems such as CLEAN \cite{CLEAN}, ECPred \cite{ECPred}, DeepEC \cite{DeepEC}, and ProteInfer \cite{ProteInfer}. Extensions including CLEAN-contact \cite{CLEAN-contact}, CREEP \cite{yang2024care}, and PhiGNet \cite{jang2024accurate} incorporate structural, chemical, or evolutionary information to further improve performance. Yet these models share a key limitation: they rely on conventional supervised training pipelines that assume the availability of large amounts of labeled data and treat all sample annotations as equally informative. With the vast majority of proteins remaining uncharacterized \cite{rappsilber2024dive, rocha2023functional}, framing annotation as a one-time batch process is misaligned with experimental practice. In reality, only a small fraction of candidate proteins can be labeled at any given time, and the information each annotation provides to a function prediction model differs across proteins.

\begin{figure*}[t]
\centering
  \includegraphics[width=0.9\linewidth]{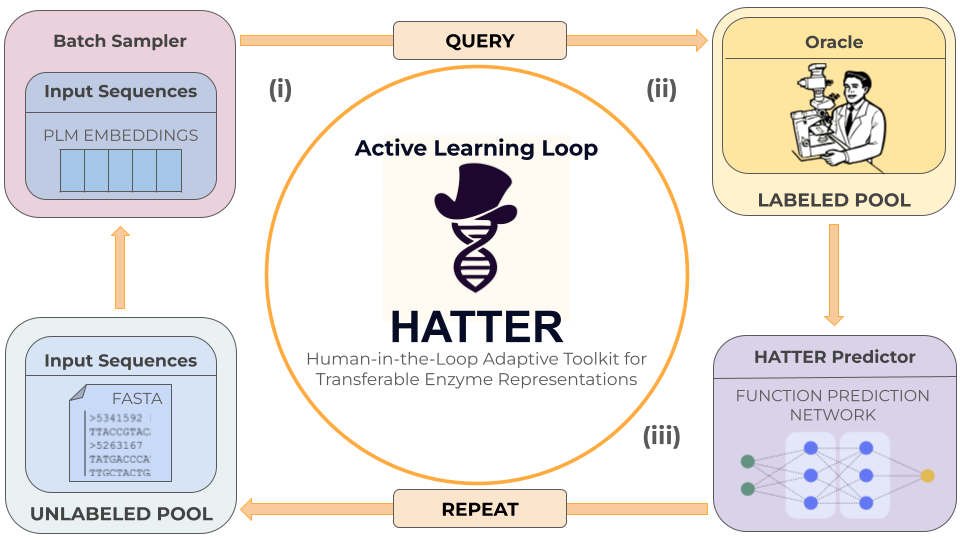}
  \caption{A graphical demonstration of the HATTER pipeline. The HATTER pipeline mirrors a standard active learning pipeline and follows four steps: (i) Select queries (amino acid sequences) based on predicted model ``uncertainty'', where uncertainty can be calculated using several implemented methods, (ii) label model queries with real-life experimental annotation (i.e. the ``oracle''), (iii) use the labeled queries to update the functional prediction model, and (iv) repeat steps (i)-(iii) until model convergence or desired performance is achieved.}
  \label{fig:AL-generic}
\end{figure*}

This disconnect highlights a deeper need: AI systems for function prediction that evolve alongside new data and the biologists who generate and interpret that data. Active learning (AL) provides a natural solution. Active learning (Figure \ref{fig:AL-generic}) is a machine learning technique that aims to improve model performance efficiently by selectively choosing the most informative data points for training on each batch from an unlabeled pool \cite{huseljicinterplay}, decreasing the amount of data and computational time needed to achieve strong training performance. Instead of randomly sampling data from a massive pool, active learning strategies identify and select examples that are expected to yield the greatest improvement in the model. 



Given the need for cheaper and faster biological labeling and experimentation, AL has been increasingly proposed as a potential solution for improving training efficiency in biological settings. AL for biology is most established in protein engineering settings, where active learning samplers can iteratively propose the next variants to test in local mutational neighborhoods coupled with high-throughput screening (\cite{yang2025, Hu2023BOEVO, Vornholt2024ArMActiveLearning, Thomas2025TeleProt, Wittmann2021TrainingSetDesign, Biswas2021LowN}). Related applications have applied AL in semi-autonomous lab environments to propose the most informative experiments to conduct that optimize a desired biological phenotype (\cite{dama2023bacterai, Coutant2019ClosedLoopYeast, Naik2016ActiveMLExperimentation}). AL has also been used for function-prediction-adjacent tasks such as efficient substrate-scope mapping (\cite{Pertusi2017SimAL}). To our knowledge, however, AL has not been applied to enzyme discovery and function prediction, in which uncharacterized enzymes to experimentally label are selected for greatest general improvement in enzyme functional annotation.


%

To enable more efficient experimental annotation and training of enzyme function prediction models, we introduce HATTER (Human-in-the-loop Adaptive Toolkit for Transferable Enzyme Representations), a first-ever framework for human-in-the-loop (HITL) enzyme function prediction. HATTER is designed explicitly for iterative experimental deployment, enabling biologists to query, annotate, and update enzyme function prediction models across sequential experimental rounds. HATTER integrates multiple AL strategies to select the most informative unannotated protein sequences with human-in-the-loop input of novel experimentally labeled proteins, enabling rapid functional prediction model updates alongside a benchmarking suite for comparing acquisition strategies. Its modular, model-agnostic design accommodates the state-of-the-art functional prediction model CLEAN \cite{CLEAN}, generic neural network classifiers, and custom architectures implemented through DAL-Toolbox \cite{huseljicinterplay}.

In addition to our public software, this work makes two contributions. First, we present a systematic evaluation of active learning for large-scale enzyme function prediction under realistic sequence- and label-stratified settings. Second, we demonstrate that active learning achieves statistically comparable predictive performance to standard supervised training while requiring substantially less training data, thereby reducing labeling costs and computational effort through faster convergence, fewer model updates, and processing less data. Together, these findings highlight the potential for update-efficient HITL workflows to accelerate protein function discovery, providing a foundation for scalable, computationally efficient AI systems that reduce experimental and computational cost while maintaining competitive predictive performance.


\section{Methodology}

HATTER is a complete active learning system for functional discovery, supporting multiple model architectures, acquisition functions, and providing a suite of experimental, deployment, and evaluation protocols (Figure~\ref{fig:AL-generic}). We describe below the HATTER implementation and the experimental protocols used to evaluate active learning in enzyme function prediction for scenarios with both overlapping and non-overlapping functional labels. 


\subsection{HATTER}

HATTER is explicitly designed with HITL considerations: (i) adjustable batch sizes that reflect real experimental throughput (as few as 1 per batch); (ii) queries exported in annotated format; (iii) \texttt{update} mode designed for experimental fine-tuning; (iv) simulation mode providing pre-experimental planning; and, (v) ease of loading/storing model weights to avoid re-training between rounds. The HATTER software is provided as a python-based toolkit with additional documentation of each mode in the associated Github repo: https://github.com/hoarfrost-Lab/HATTER. 

\paragraph{Inputs:}

Input data are provided as tab-separated files containing UniProt protein identifiers, Enzyme Commission (EC) numbers representing functional labels, and amino acid sequences. Training datasets contain EC labels, while pool datasets represent unlabeled or partially labeled candidate sequences and therefore the EC column initially begins blank. Validation and test datasets are used to compute evaluation metrics. 


\paragraph{Underlying Model:}

HATTER uses the CLEAN functional prediction model \cite{CLEAN} as the default architecture due to its strong performance, its reliance on sequence information alone, and ease of integration with active learning. CLEAN consists of an ESM-1b encoder \cite{esm1b}, a decoder trained with triplet contrastive loss \cite{ge2018deep} to map inputs to a functional embedding space, and an inference procedure which uses pairwise distance-mapping of queries to their nearest EC cluster in this embedding space. HATTER further supports a simple two-layer neural network (referred to as ``TwoLayer'') trained under a standard supervised classification objective (cross entropy). Alternative protein language models (pLMs) can be used by adjusting the input size of the classifier.


\paragraph{Active Learning Implementation:}

We integrate active learning sampling with the underlying models by wrapping the neural network portion of each model with an AL sampler class object using the python package DAL-Toolbox (Deep Active Learning Toolbox) \cite{huseljicinterplay}. We test multiple acquisition strategies: uncertainty sampling (implemented using least confidence), entropy sampling, margin sampling, bayesian sampling (implemented as least confidence), BALD (Bayesian Active Learning by Disagreement, \cite{houlsby2011bayesian}), and query-by-committee (QBC, \cite{burbidge2007active}). The QBC learner is custom implemented as a combination of an uncertainty sampler, margin sampler, and entropy sampler, where the voting procedure and final score is calculated as the maximum disagreement of the committee. We note that for the bayesian sampling procedure, we replace dropout layers of the original underlying network with Monte Carlo Dropout (MCDropout) layers to satisfy the requirements of using bayesian sampling in DAL-Toolbox \cite{huseljicinterplay}. We additionally implement a ``random'' sampler as a baseline which utilizes the same uncertainty sampler in the DAL-Toolbox package, but is adapted to query data randomly by calculating a random score from a normal distribution between 0-1 for each potential data input and using these random scores to perform the query. 

\paragraph{Operational Modes and Outputs:}

Our software implements four unique modes designed to create a human-in-the-loop compatible pipeline:

\begin{enumerate}
    \item \textbf{Training Mode (train).}
    Training mode performs standalone pretraining of the selected function prediction model. This mode is used to produce model checkpoints from labeled data, which can subsequently be reused later in \texttt{init} or \texttt{update} modes. Training metrics and optional validation results, as well as model weight checkpoints, are saved as outputs. Training can be done using both standard or active learning paradigms and various parameters (e.g. early stopping, learning rate, etc.) can be specified via command line arguments.

    \item \textbf{Initialization Mode (init).}
    Initialization mode performs the first active learning round. This is to ``seed'' the active learning as there are no initially labeled points in the unlabeled pool. 
    A fixed number of candidate sequences are selected according to the specified batch size and query strategy parameters before being written to an output file for experimental evaluation. 
    
    \item \textbf{Update Mode (update).}
    Update mode incorporates experimental results from a completed active learning round. This assumes the output query file is passed from the previous round (either \texttt{init} or \texttt{update\_w\_requery}) with the ``Result'' column filled in. The model is fine tuned on these newly labeled sequences, and these sequences are removed from the pool and merged into the training set. 

    \item \textbf{Simulation Mode (simulation).}
    Simulation mode emulates multiple rounds of active learning (\texttt{train} [optional] + \texttt{init} + multiple \texttt{update} rounds) without requiring ``real'' HITL experimental feedback. Labeled datasets are treated as oracle annotations, allowing the pipeline to iteratively select, label, and update models \emph{in silico}. This mode is used to evaluate active learning strategies, batch sizes, and hyperparameters, and to assess expected performance trends prior to committing to experimental data collection.

\end{enumerate}


\paragraph{Evaluation Metrics:}

HATTER automatically computes the following metrics: Area Under the Receiver Operator Curve (AUROC) and Area Under the Precision Recall Curve (AUPRC, more immune to class imbalance \cite{mcdermott2024closer}), Accuracy, Precision (false positive rate), Recall (false negative rate), F1-score (harmonic mean of precision and recall) and hierarchical F1-score (hF1), which is a version of F1-score which accounts for precision/recall at various levels of a hierarchical label (important for EC scores). The hierarchical metrics are computed using the hiclass \cite{miranda2023hiclass} python package and can be ignored for non-hierarchical classification.

\subsection{Active Learning Benchmarking using HATTER}

In order to demonstrate the utility of HATTER, we perform a controlled benchmarking of active learning strategies and assessment of model behavior using datasets of varying enzyme prediction difficulty. 

\paragraph{Data:}

We use a genetically stratified 5-fold cross-validation dataset \cite{GRIMM}, where each cross-validation split is stratified respective of both UniRef50 cluster ID and EC number to maintain maximum sequence divergence between training and evaluation sets for each functional label. A training set and validation set are defined for each split, along with two test sets: \texttt{Test-1}, which follows the same EC label distribution as the training and validation sets, and \texttt{Test-2}, a held-out test set containing EC numbers not present in the train, validation, or \texttt{Test-1} partitions of that split \cite{GRIMM}.
This dataset is carefully selected due to the fact that it captures biological novelty in both the feature space (e.g. sequence diversity) and label space (EC label representation). We further evaluated performance on three additional held-out datasets---\texttt{price-129}, \texttt{new-392}, and \texttt{halogenase}---originally used by CLEAN \cite{CLEAN}, which include EC numbers not fully overlapping with the training data.

\paragraph{Training Procedure:}

To perform the active learning training, we train each of HATTER's underlying models (CLEAN and TwoLayer) from scratch, for each acquisition strategy, for a maximum of 100 epochs, using an Adam optimizer and learning rate of 5e-4. We perform early stopping at the epoch of lowest validation loss across the 100 epochs of training. We note that unlike the original CLEAN, we do not implement the adaptive rate procedure when training our models due to a reduced number of total epochs. Further, because the cross-validation splits are not truly data independent (due to the UniRef50 clustering overlap \cite{GRIMM}), we instead train five \textit{independent} models (one per split) to avoid data overlap and compute ensemble metrics/confidence intervals post-training. We report evaluation metrics with 95\% confidence intervals over the five splits.

Standard training is performed for both underlying models using the same parameters (100 epochs, Adam optimizer, learning rate 5e-4, with early stopping), but with batches determined by the default pytorch DataLoader (batch\_size = 6000, shuffle = True). 

\paragraph{Implementation Details:}

All experiments were performed using Python 3.11 and CUDA 12.1. Models were trained on a single NVIDIA A100 (80 GB) GPU, with typical training times under 24 hours depending on batch size and sequence length. We note that experiments reported here were conducted using \texttt{train} mode unless otherwise specified.

\begin{table*}[t]
\centering
\small
\caption{Combined F1-score table for CLEAN and TwoLayer models. Delta values in parentheses indicate change vs. standard training. Bold numbers indicate the best performance per column; underline represents second best. Colormap follows performance values with orange representing higher performance shading into white representing lowest performance.}
\resizebox{\textwidth}{1.5cm}{
\footnotesize
\begin{tabular}{l|ccccc|ccccc}
\toprule
& \multicolumn{5}{c|}{\textbf{CLEAN (F1-score)}} 
& \multicolumn{5}{c}{\textbf{TwoLayer NN (F1-score)}} \\
\textbf{Active Learner} 
& Test-1 & Test-2 & new-392 & price-129 & halogenase
& Test-1 & Test-2 & new-392 & price-129 & halogenase \\
\midrule
standard            
& \shadecell{shade17}{0.812 (0.000)} & \shadecell{shade12}{\textbf{0.596 (0.000)}} & \shadecell{shade10}{0.524 (0.000)} & \shadecell{shade4}{\textbf{0.295 (0.000)}} & \shadecell{shade10}{\textbf{0.509 (0.000)}}
& \shadecell{shade15}{\textbf{0.775 (0.000)}} & \shadecell{shade10}{\textbf{0.496 (0.000)}} & \shadecell{shade10}{\textbf{0.534 (0.000)}} & \shadecell{shade4}{\textbf{0.299 (0.000)}} & \shadecell{shade9}{\textbf{0.437 (0.000)}} \\

random\_sampling    
& \shadecell{shade20}{\textbf{0.843 (+0.031)}} & \shadecell{shade12}{0.590 (-0.006)} & \shadecell{shade14}{\textbf{0.575 (+0.051)}} & \shadecell{shade3}{0.279 (-0.016)} & \shadecell{shade9}{0.462 (-0.047)}
& \shadecell{shade14}{0.756 (-0.019)} & \shadecell{shade8}{0.410 (-0.086)} & \shadecell{shade10}{0.486 (-0.048)} & \shadecell{shade3}{0.257 (-0.042)} & \shadecell{shade5}{0.316 (-0.121)} \\

uncertainty\_sampling
& \shadecell{shade20}{\underline{0.842 (+0.030)}} & \shadecell{shade12}{0.592 (-0.004)} & \shadecell{shade14}{\underline{0.568 (+0.044)}} & \shadecell{shade2}{0.250 (-0.045)} & \shadecell{shade9}{0.466 (-0.043)}
& \shadecell{shade15}{\underline{0.767 (-0.008)}} & \shadecell{shade11}{\underline{0.485 (-0.011)}} & \shadecell{shade10}{\underline{0.525 (-0.009)}} & \shadecell{shade3}{\underline{0.287 (-0.012)}} & \shadecell{shade8}{\underline{0.403 (-0.034)}} \\

entropy\_sampling   
& \shadecell{shade20}{0.841 (+0.029)} & \shadecell{shade12}{\underline{0.593 (-0.003)}} & \shadecell{shade13}{0.560 (+0.036)} & \shadecell{shade3}{\underline{0.283 (-0.012)}} & \shadecell{shade8}{0.433 (-0.076)}
& \shadecell{shade12}{0.725 (-0.050)} & \shadecell{shade8}{0.452 (-0.044)} & \shadecell{shade10}{0.518 (-0.016)} & \shadecell{shade3}{0.255 (-0.044)} & \shadecell{shade6}{0.336 (-0.101)} \\

margin\_sampling    
& \shadecell{shade20}{0.841 (+0.029)} & \shadecell{shade12}{0.590 (-0.006)} & \shadecell{shade13}{0.556 (+0.032)} & \shadecell{shade3}{0.279 (-0.016)} & \shadecell{shade9}{\underline{0.473 (-0.036)}}
& \shadecell{shade15}{\underline{0.767 (-0.008)}} & \shadecell{shade11}{0.482 (-0.014)} & \shadecell{shade10}{0.522 (-0.012)} & \shadecell{shade3}{0.279 (-0.020)} & \shadecell{shade8}{0.392 (-0.045)} \\

QBC                 
& \shadecell{shade20}{0.840 (+0.028)} & \shadecell{shade12}{0.590 (-0.006)} & \shadecell{shade13}{0.559 (+0.035)} & \shadecell{shade3}{0.266 (-0.029)} & \shadecell{shade9}{0.457 (-0.052)}
& \shadecell{shade11}{0.694 (-0.081)} & \shadecell{shade0}{0.196 (-0.300)} & \shadecell{shade7}{0.387 (-0.147)} & \shadecell{shade1}{0.181 (-0.118)} & \shadecell{shade0}{0.091 (-0.346)} \\

bayesian            
& \shadecell{shade17}{0.813 (+0.001)} & \shadecell{shade7}{0.457 (-0.139)} & \shadecell{shade11}{0.526 (+0.002)} & \shadecell{shade3}{0.281 (-0.014)} & \shadecell{shade6}{0.354 (-0.155)}
& \shadecell{shade14}{0.760 (-0.015)} & \shadecell{shade9}{0.437 (-0.059)} & \shadecell{shade10}{0.503 (-0.031)} & \shadecell{shade3}{0.228 (-0.071)} & \shadecell{shade6}{0.339 (-0.098)} \\

BALD                
& \shadecell{shade17}{0.812 (0.000)} & \shadecell{shade7}{0.473 (-0.123)} & \shadecell{shade11}{0.529 (+0.005)} & \shadecell{shade3}{0.256 (-0.039)} & \shadecell{shade6}{0.344 (-0.165)}
& \shadecell{shade14}{0.759 (-0.016)} & \shadecell{shade9}{0.437 (-0.059)} & \shadecell{shade10}{0.501 (-0.033)} & \shadecell{shade3}{0.228 (-0.071)} & \shadecell{shade5}{0.314 (-0.123)} \\

\bottomrule
\end{tabular}
}
\label{fig:performance-pretraining}
\end{table*}

\section{Results}

\subsection{Active Learners perform comparatively to standard training procedure}

We systematically compared active learning to standard supervised training across multiple acquisition strategies, model architectures, and biologically stratified evaluation datasets (Table \ref{fig:performance-pretraining}). 
Standard training generally achieved the highest absolute F1-scores; however, the performance gap relative to active learning was consistently small, on average -0.042 (-0.019 for CLEAN and -0.066 for TwoLayer), despite the limited training time for active learning training. Active learning slightly \textit{improved} performance relative to standard training in two instances, for the CLEAN model with uncertainty sampling on the \texttt{Test-1} and \texttt{new-392} sets, achieving a score of 0.842 on \texttt{Test-1} (+0.030 improvement to standard training at 0.812) and 0.568 on \texttt{new-392} (+0.044 improvement to standard training at 0.524). 

CLEAN performs best on the \texttt{Test-1} set relative to other evaluation datasets, which has the greatest similarity to training data \cite{GRIMM}. For the remaining evaluation datasets, performance was lower and more variable. On \texttt{Test-2}, standard training on CLEAN achieved the highest F1 of 0.596, while uncertainty and entropy sampling were slightly lower at 0.592 (-0.004 delta) and 0.593 (-0.003 delta). For the \texttt{price-129} and halogenase datasets, standard training on CLEAN remained the best-performing approach, achieving 0.295 and 0.509, respectively, while AL strategies yielded lower F1-scores (e.g., entropy sampling: 0.283 and 0.433 respectively). Bayesian-based samplers consistently underperformed relative to point-estimate acquisition strategies and standard training in most cases (\texttt{Test-1}, \texttt{Test-2}, \texttt{new-392}, and \texttt{halogenase}); however, bayesian sampling performed third best on \texttt{price-129} behind entropy sampling and standard training.

For the TwoLayer classifier, standard training consistently outperformed all AL strategies across all test sets. For example, on \texttt{Test-1} standard training achieved 0.775 compared to 0.756 (-0.019 delta) for random sampling and 0.767 (-0.008 delta) for uncertainty sampling, while on \texttt{halogenase} standard training achieved 0.437 while random sampling only reached 0.316 (-0.121 delta). TwoLayer performance was consistently lower than CLEAN for all training strategies, despite similar model sizes.

\begin{figure*}[h]
\centering
  \includegraphics[width=1.0\linewidth]{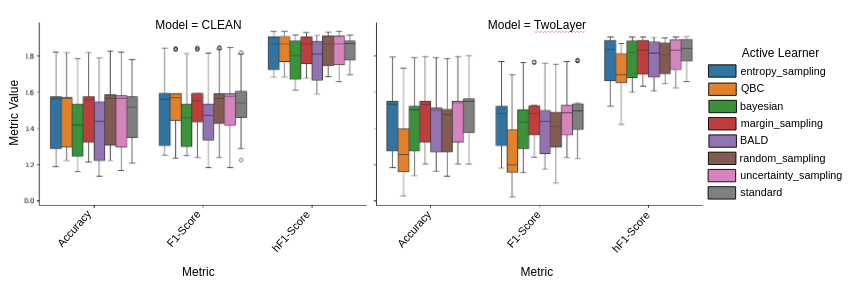}
  \caption{A comparison of performance for both underlying models (left: CLEAN, right: TwoLayer) across all acquisition strategies. Across accuracy, F1, and hF1 metrics, there is no statistically significant difference among acquisition functions, with all acquisition functions within the 95\% confidence interval for each other as well as random sampling and standard training.}
  \label{fig:pre-AL-boxplot}
\end{figure*}

Overall, across all models, acquisition functions, and datasets, performance under active learning is comparable to standard training, with modest improvements observed in specific cases (e.g., CLEAN on \texttt{Test-1} and \texttt{new-392}). No single AL strategy consistently outperforms standard training across all evaluation settings, although point-estimate uncertainty sampling strategies (uncertainty, entropy, margin sampling and QBC) generally outperform bayesian methods (bayesian sampling, BALD). Notably, across all datasets and acquisition strategies, active learning achieved performance that was statistically indistinguishable from standard training within 95\% confidence intervals (Figure \ref{fig:pre-AL-boxplot}), despite processing substantially less data and requiring fewer model updates.

\subsection{Active learning matches standard training performance with far fewer model updates}

\begin{figure}[t]
\centering

\begin{minipage}{0.49\textwidth}
    \centering
    \includegraphics[width=1.0\linewidth]{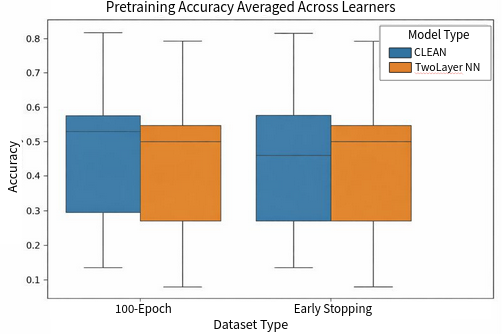}
\end{minipage}
\hfill
\centering
\begin{minipage}{0.49\textwidth}
    \centering
    \resizebox{\textwidth}{!}{
    \begin{tabular}{lrr}
    \toprule
    Strategy & Avg. Epoch & 95\% Conf. \\
    \midrule
    standard & 86.2 & 8.76 \\
    random\_sampling & 67.00 & 28.83 \\
    uncertainty\_sampling & 81.20 & 15.21 \\
    entropy\_sampling & 58.80 & 22.54 \\
    margin\_sampling & 72.40 & 16.13 \\
    QBC & 74.60 & 16.15 \\
    bayesian & 78.60 & 15.13 \\
    BALD & 79.40 & 9.95 \\
    \bottomrule
    \end{tabular}
    }
\end{minipage}

\caption{Early stopping analysis. \textbf{Left:} Distribution of model accuracy for CLEAN (blue) and TwoLayer (orange) across five-fold cross validation splits comparing early stopping (right) and full 100 epoch (left) training. \textbf{Left:} Average epoch and 95\% confidence interval across five-fold splits where early stopping occurred for each strategy using CLEAN.}
\label{fig:combined-early-stopping}
\end{figure}

We next compared the number of epochs required to reach peak performance during standard and active learning–based training with early stopping (Figure \ref{fig:combined-early-stopping}). Across all acquisition strategies as well as standard training, early stopping resulted in comparable performance to training for the full 100 epochs, with no statistically significant differences observed between early-stopped models and fully trained models. This suggests that early stopping did not meaningfully degrade model performance for either standard or active learning–based training.

While final performance was comparable across paradigms, we observed notable differences in the number of epochs required to reach peak performance. Standard training converged most slowly, with an average early stopping point of 86.2 epochs, and exhibited the lowest variability across folds (± 8.76, 95\% confidence). In contrast, active learning strategies generally reached early stopping in fewer epochs, averaging approximately 70 epochs, but with substantially higher variability depending on the acquisition function. 


Among the active learners, entropy sampling exhibited the earliest average stopping point at 58.8 (± 22.54) epochs, while random sampling converged the second earliest at 67.0 (± 28.83) epochs. However, these strategies also showed the largest confidence intervals, indicating less consistent convergence behavior across folds. Other acquisition functions, including margin sampling (72.4 ± 16.13), QBC (74.6 ± 16.15), Bayesian sampling (78.6 ± 15.13), and BALD (79.4 ± 9.95), fell between these extremes, converging earlier and with greater variability than standard training but later and with lower variability than entropy sampling.

Overall, these results indicate that active learning can reach comparable performance to standard training with fewer model updates on average, albeit with greater variability across strategies and folds. Training to the full 100 epochs yielded slightly higher performance in some cases, but these gains were not outside the 95\% confidence intervals, suggesting diminishing returns beyond the early stopping point (Figure~\ref{fig:combined-early-stopping}).




\subsection{Data Efficiency and Computational Speedup in Active Learning}


\begin{figure}[t]
\centering

\begin{minipage}{0.5\textwidth}
    \centering
    \includegraphics[width=1.0\linewidth]{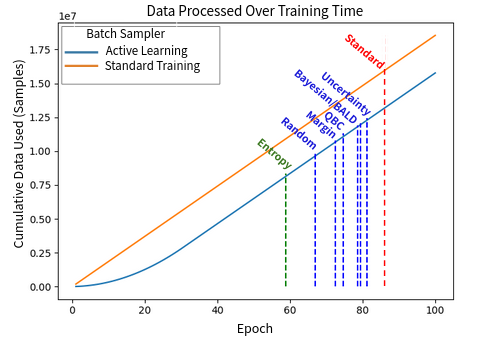}
\end{minipage}
\hfill
\centering
\begin{minipage}{0.45\textwidth}
    \centering
    \resizebox{\textwidth}{!}{
    \begin{tabular}{lr}
    \toprule
    Strategy & \% Data Reduced \\
    \midrule
    standard & 0\% \\
    random\_sampling & 39.5\% \\
    uncertainty\_sampling & 23.0\% \\
    entropy\_sampling & 48.0\% \\
    margin\_sampling & 33.2\% \\
    QBC & 30.6\% \\
    bayesian & 26.0\%\\
    BALD & 25.1\% \\
    \bottomrule
    \end{tabular}
    }
\end{minipage}

\caption{Data efficiency and computation analysis. \textbf{Left:} Comparison of data usage and computational efficiency between active learning and standard training, showing cumulative data processed over 100 epochs. Dotted vertical lines represent the epoch of early stopping for the fastest-converging AL strategy (entropy sampling, green) vs. standard training (red). Blue vertical lines indicate early stopping point for other sampling strategies. \textbf{Right:} Percent data processing reduction calculated at the epoch of early stopping relative to standard training.}
\label{fig:speedup-comparison}
\end{figure}

It is important to note that the definition of an ``epoch'' differs between paradigms, which further widens the data efficiency savings when paired with the early stopping results (Figure~\ref{fig:speedup-comparison}). In standard training, an epoch processes the full 185,418-sample dataset, whereas in active learning it corresponds to a single model update from a selected batch from the pool added to the previously labeled pool, in this example 6,000 samples in epoch 0, and increasing in increments of 6,000 per epoch until all data is seen at epoch 31. Because active learning processes substantially fewer samples per model update than standard training, we quantify training efficiency in terms of total samples processed rather than epochs alone. It follows that with fewer epochs required for active learning convergence (Figure~\ref{fig:combined-early-stopping}), and fewer samples processed per epoch, this results in lower data processing and computation required to achieve comparable performance. 



Under standard training, the full dataset is processed repeatedly each epoch (in 31 batches per epoch), resulting in 3,100 batch updates (or 18,600,000 samples processed) over the full training budget of 100 epochs. By contrast, active learning incrementally incorporates 6,000 newly acquired samples in a single batch per training step into the labeled pool. Due to the size of the training set and the constraint that samples cannot be resampled once added to training, the full dataset is covered after approximately 30 epochs and subsequent epochs use 185,418 samples per epoch, resulting in 15,949,260 samples processed in total over the full 100 epochs (Figure~\ref{fig:speedup-comparison})---a roughly 14.5\% reduction. 

Importantly, active learning has already reached comparable performance to the maximum performance by the point of early stopping (Figure~\ref{fig:combined-early-stopping}), despite having processed substantially less data overall. At the early stopping point for the fastest-converging active learning strategy (59 epochs, entropy sampling, Figure~\ref{fig:combined-early-stopping}), only 8,347,122 samples were processed compared to 15,945,948 samples for standard learning with early stopping at 86 epochs (Figure \ref{fig:speedup-comparison}), resulting in a ~48.0\% reduction in data processing time. Even the uncertainty\_sampling---which has the longest training time of all AL samplers---sees a ~23.0\% reduction in overall data for comparable performance. 

Coupled with the early stopping results, these findings demonstrate that active learning is substantially more data-efficient than standard training: it reaches comparable performance after processing far less data and with significantly fewer batch computations, while ultimately achieving full dataset coverage when needed.





\section{Discussion}

Our results collectively demonstrate that active learning strategies perform comparatively to standard training while offering substantial computational advantages, indicating that active learning could be used to reduce the experimental effort of functional annotation required to improve state-of-the-art function prediction models. We find that while no acquisition function consistently outperforms standard training (Table \ref{fig:performance-pretraining}), the performance of standard versus active learning trained models is statistically indistinguishable (Figure \ref{fig:pre-AL-boxplot}). This is the case both where active learning outperforms standard training -- e.g. CLEAN on \texttt{Test-1} -- and also where standard training outperforms active learning -- e.g. \texttt{Test-2}, \texttt{new-392}, and \texttt{price-129}. In practical enzyme discovery settings, where experimental validation capacity constrains the number of sequences that can be annotated per iteration, such reductions in data processing and update frequency may translate directly into shorter experimental cycles and reduced costs for both computational and experimental effort.



We further find that active learners generally converge faster than standard training, requiring approximately 70 epochs on average (59 best case) versus 86 for standard training, while achieving comparable performance (Figure \ref{fig:combined-early-stopping}). There is a consistent correlation between the highest performing acquisition functions (random and entropy sampling, Table \ref{fig:performance-pretraining}) and the earliest epoch of stopping. This trend suggests that these acquisition strategies may identify more informative samples more quickly than other AL strategies (Figure \ref{fig:speedup-comparison}), and further supports the finding that active learning can converge on a performant functional prediction model more quickly than standard training. 

Beyond performance, active learning offers clear advantages in data efficiency and computational cost (Figure \ref{fig:speedup-comparison}). By incrementally adding only one batch of samples to the labeled pool (in our experiments, 6000 samples) per model update instead of using the full dataset, active learning achieves roughly a 14.25\% reduction in total batch computations over the full 100 epochs, and up to 49.0\% when compared at the point of early stopping. Although with our small training data (185,418 samples) we often process the full dataset within the early stopping period, for large datasets (~1,000,000+ sequences), this approach is expected to produce models that reach near-equivalent performance with substantially less data processed per epoch compared to standard supervised training, enabling faster experimentation and reduced computational overhead, without sacrificing model quality \cite{huseljicinterplay}.

Of considerable interest throughout our experiments is that random sampling appears to perform equal to or in some cases better than the other acquisition functions, and typically differs from standard training, a behavior that has been previously observed  \cite{huseljicinterplay, gashi2024deep}. We suggest two potential explanations for this outcome: First, the random sampler manages to capture more input/sequence diversity simply by the fact that it is not optimizing for an arbitrary measure of embedding uncertainty, as with the other acquisition functions. Sequence similarity and biodiversity are known to heavily affect model performance for this particular task \cite{CLEAN, GRIMM, yang2024care}, so in a scenario where embedding uncertainty does not align with sequence similarity such sampling strategies may adversely affect the model. Notably, the \texttt{halogenase} dataset remains particularly challenging, exhibiting the lowest performance across all training strategies, which coincides with the lowest sequence similarity overlap with training data \cite{GRIMM}. This relationship is consistent with prior observations that in settings with highly diverse input data random sampling and standard training can perform comparably to active learning approaches \cite{gashi2024deep}.


Second, deep active learning is known to struggle with estimating uncertainty in very large embeddings \cite{huseljicinterplay}. Our observation that random sampling performs comparably to uncertainty-based acquisition functions aligns with prior work \cite{gashi2024deep, zheng2002active, beyer2015select, huseljicinterplay}, which demonstrates that deep active learning does not reliably improve performance compared to standard training in complex, high-dimensional tasks. We see in particular that active learning does not pair well with the TwoLayer classifier, which due to the nature of the 5000+ EC classes has a very large final layer compared to CLEAN's 256 nodes. The contrastive learning objective of CLEAN may further alleviate the dimensionality challenge of this task, as contrastive learning is well suited to separating classes in embedding space when labels are imbalanced or sparse. This difference in performance between TwoLayer and CLEAN may therefore be an indication that the complexity and dimensionality of the functional classification task is a challenge for the active learning setting - particularly with the relatively small available training data for this task. While we cannot pinpoint the exact cause of this performance from our experiments, we theorize both of these issues (input/sequence diversity + output/embedding size) need to be addressed in future iterations of biological active learning efforts. 


Altogether these findings suggest that for enzyme function prediction, the primary advantage of active learning lies not in surpassing fully supervised training in static benchmarks, but in enabling adaptive, iterative model improvement aligned with experimental constraints. In large-scale enzyme discovery efforts, where experimental validation capacity is the limiting factor, reducing the number of model updates and labeled samples required to reach competitive performance may be more valuable than marginal improvements in peak accuracy. In real-world HITL deployments, model updates would occur after each experimental round, often involving tens-to-hundreds rather than thousands of sequences. Our simulation framework therefore provides an upper-bound estimate of achievable efficiency under simulated, large-batch conditions. In many enzyme discovery settings, only tens to hundreds of candidates can be experimentally validated per round. HATTER’s batch-based update mechanism is explicitly designed to operate within such constraints, supporting iterative refinement without full retraining. Future work will investigate smaller-batch, experimentally realistic regimes to better characterize performance in practical settings and to assess how update frequency and limited data per round affect overall learning dynamics.


\section{Conclusion}

We present HATTER, a modular human-in-the-loop active learning framework designed to improve the data efficiency and generalizability of deep learning models for enzyme functional prediction. By providing a modular active learning framework, HATTER bridges the gap between static benchmark evaluation and real-world human-in-the-loop enzyme discovery, where models must continuously evolve alongside newly generated experimental data. We demonstrate that active learning within HATTER performs competitively with standard supervised training while achieving greater data efficiency and faster convergence. Entropy and uncertainty-based acquisition strategies show particular promise for rapidly improving model accuracy and precision with less data. Future work will integrate HATTER’s modular architecture with test-time adaptation, enabling human-in-the-loop fine-tuning during both out-of-distribution inference and de novo experimental tasks. This approach has the potential to create biological AI tools that are not only predictive but adaptive and collaborative, capable of incorporating expert insight to adapt to novel biology and accelerate functional discovery in uncharted regions of sequence space.

\section*{Acknowledgment}

This research was developed with funding from the Defense Advanced Research Projects Agency (DARPA) under the DUF Advanced Research Concept. The views, opinions and/or findings expressed are those of the authors and should not be interpreted as representing the official views or policies of the Department of Defense or the U.S. Government.

\bibliographystyle{unsrtnat}
\bibliography{references}  

\end{document}